\documentclass[aps,twocolumn,nofootinbib, superscriptaddress, floatfix]{revtex4}
\usepackage{graphicx} 
\usepackage{dcolumn} 
\usepackage{bm} 
\usepackage{amssymb} 
\usepackage{amsmath} 

\usepackage{color}

\usepackage{placeins}

\usepackage{hyperref}

\definecolor{darkgreen}{RGB}{50,150,0}

\newcommand{\mdp}{m_{\gamma'}}
\newcommand{\mueV}{\mu\mathrm{eV}}
\newcommand{\eV}{\ \mathrm{eV}}
\newcommand{\GeV}{\ \mathrm{GeV}}

\usepackage{xcolor}
\usepackage{bm}
\usepackage{cleveref}

\usepackage{chngcntr}
\preprint{TU-1074,IPMU18-015,RESCEU-13/18,MIT-CTP/5066}

\begin{document}

\title{Relic Abundance of Dark Photon Dark Matter}
\author{Prateek Agrawal}
\affiliation{Jefferson Physical Laboratory, Harvard University,
17 Oxford Street, Cambridge, MA 02138, USA}
\author{Naoya Kitajima}
\affiliation{Department of Physics and Astrophysics, Nagoya University, Chikusa, Nagoya 464-8602, Japan}
\author{Matthew Reece}
\affiliation{Jefferson Physical Laboratory, Harvard University,
17 Oxford Street, Cambridge, MA 02138, USA}
\author{Toyokazu Sekiguchi}
\affiliation{Research Center for the Early Universe (RESCEU), Graduate School of Science, the University of Tokyo, Tokyo 113-0033, Japan}
\author{Fuminobu Takahashi}
\affiliation{Department of Physics, Tohoku University, Sendai, Miyagi 980-8578, Japan}
\affiliation{Kavli Institute for the Physics and Mathematics of the Universe (Kavli IPMU), UTIAS, WPI, The University of Tokyo, Kashiwa, Chiba 277-8568, Japan}
\affiliation{Center for Theoretical Physics, Massachusetts Institute of Technology, Cambridge, MA 02139, USA}
\date{\today}

\begin{abstract}
We present a new mechanism for producing the correct relic
abundance of dark photon dark matter over a wide range of its mass,
extending down to
$10^{-20}\eV$.
The dark matter abundance is initially stored in an axion which is
misaligned from its minimum. When the axion starts oscillating, it
efficiently transfers its energy into dark photons via a tachyonic
instability.
If the dark photon mass is within a few orders of magnitude of the axion mass, 
$m_{\gamma'}/m_a = {\cal O}(10^{-3} - 1)$, 
then dark photons make up the
dominant form of dark matter today. We present a numerical lattice
simulation for a benchmark model that explicitly realizes our
mechanism. This mechanism firms up the
motivation for a number of experiments searching for dark photon dark
matter. 
\end{abstract}

\maketitle

\section{Introduction}
An intriguing possibility is that dark matter (DM) is made up of very light
spin-1 bosons. This would belie our usual intuition that light spin-1
bosons mediate forces but do not make up matter in our universe. Very
light bosonic DM ($m \lesssim 1~{\rm eV}$) has interesting properties.
It has high occupation numbers within galaxies (the de Broglie
wavelength of the particles is larger than the interparticle spacing)
\cite{Baldeschi:1983mq}, so it can be modeled as a classical field. In
the extremely low mass range, spin-1 bosonic DM could give a
realization of the Fuzzy Dark Matter paradigm \cite{Hu:2000ke,
Hui:2016ltb}, with many associated effects on galactic structure. If
DM in our universe is a very light boson, a rich variety of dark
condensed matter physics could await discovery.

A long-standing problem for dark photon DM models is to obtain the correct relic
abundance of DM. It is possible to get a relic abundance
from inflationary fluctuations~\cite{Graham:2015rva}. This is estimated to be
\begin{align}
  \Omega_{\gamma'}
  =
  \Omega_{DM}
  \sqrt{\frac{\mdp}{6\times 10^{-6}\eV}}
  \left( \frac{H_I}{10^{14}\GeV}\right)^2,
\end{align}
where $\Omega$ is the energy density relative to the critical density
today, and $H_I$, the inflationary Hubble scale, is bounded to be $\lesssim
10^{14}\GeV$ from the lack of observation of primordial tensor modes.
Hence, dark photons from inflationary fluctuations can only constitute
all of the DM for $\mdp \gtrsim \mu{\rm eV}$. 
A misalignment
mechanism like the axion~\cite{Nelson:2011sf} may work
in a wider mass range, but it requires additional
large, highly tuned couplings to the curvature
$\mathcal{R}$~\cite{Arias:2012az}. However, by themselves, these 
couplings lead to perturbative
unitarity violation at low energies in longitudinal photon--graviton
scattering.


In this letter we present a new mechanism to populate dark photon dark
matter over a wide range of masses, extending down to $10^{-20}\eV$.
The DM abundance is initially stored in a misaligned axion,
which is coupled to the dark photon.
As the axion starts oscillating, it can efficiently transfer its
energy to the dark photons through a tachyonic instability. 
The instability persists for massive dark photons with masses
comparable to the axion. The dark photons are produced with a typical
energy close to the mass of the axion and redshift quickly to behave
as cold DM.

This instability has previously been used in many contexts: 
generation of primordial magnetic
fields
\cite{Turner:1987bw,Garretson:1992vt,Adshead:2016iae,Choi:2018dqr};
dissipation allowing for inflation in steep potentials
\cite{Anber:2009ua, Notari:2016npn, Ferreira:2017lnd}; baryo- or
leptogenesis \cite{Alexander:2004us}; 
production of chiral gravitational waves during inflation
\cite{Lue:1998mq,Alexander:2004wk,Anber:2012du,
Adshead:2013qp,Maleknejad:2016qjz,Obata:2016xcr,Dimastrogiovanni:2016fuu};
preheating \cite{Adshead:2015pva}; decreasing the abundance of QCD
axion DM \cite{Agrawal:2017eqm, Kitajima:2017peg} and
providing alternative dissipation mechanisms~\cite{Hook:2016mqo,Choi:2016kke,Tangarife:2017vnd, Tangarife:2017rgl}
in relaxion cosmology~\cite{Graham:2015cka}.

We briefly describe the mechanism for dark photon production below.
To accurately capture the production and backscattering effects of
the dark photon, a lattice numerical simulation is performed. We
describe the results of the numerical simulation. Finally, we present
two example benchmarks  which are being actively pursued in
current and planned experiments. These form ideal targets for our
mechanism, since their abundance is extremely hard to generate by
inflationary perturbations. 

In supplementary material, we discuss model building restrictions and
avenues for generating the requisite coupling of dark photons with the
axion. We also discuss models which generate observable couplings of
the dark photon to the standard model (SM).
These models are intended to be proofs of principle, and our
mechanism applies to a wide class of models. We also provide
additional details about the numerical simulation, and briefly discuss
another production mechanism where dark photons are produced
by the axion decay. 

\section{Mechanism for dark photon production}
We consider the following action of a system with the axion $\phi$ and dark photon $A_\mu$,
\begin{align}
S &= \int d^4 x \sqrt{-g} \left(
\frac{1}{2} \partial_\mu \phi \partial^\mu \phi - V(\phi) - \frac{1}{4} F_{\mu \nu} F^{\mu \nu}\right. \nonumber \\
&~~~~\left.+\frac{1}{2} m_{\gamma '}^2 A_\mu A^\mu - \frac{\beta}{4 f_a} \phi F_{\mu \nu} \widetilde{F}^{\mu \nu}\right),
\label{eq:agg}
\end{align}
with $F_{\mu \nu} = \partial_\mu A_\nu - \partial_\nu A_\mu$ and $\widetilde{F}^{\mu \nu} = \epsilon^{\mu\nu\rho\sigma} F_{\rho \sigma}/2\sqrt{-g}$ the field strength tensor and its dual, $m_{\gamma '}$  the dark photon mass, 
and $f_a$ the axion decay constant. 
The axion potential is given by
$V(\phi) = m_a^2 f_a^2 \left(1 - \cos\left(\phi/f_a\right)\right).$
We assume for our simulations that the mass of the axion is constant,
but our mechanism can be plausibly extended to a temperature-dependent
axion mass, as in the case of the QCD axion.
We denote the gauge coupling and the fine-structure constant as $g_D$ and $\alpha_D \equiv g_D^2/4\pi$, respectively. 
Here we adopt the convention $\epsilon^{0123} = 1$, $g_{\mu \nu} = (+,-,-,-)$, and  
$g \equiv {\rm det}[g_{\mu \nu}]$. The dynamical degrees of freedom are $\phi$ and ${\bf A} = \left\{A_i\right\}$, and their equations of motion
in a flat, isotropic, and homogenous universe are given by
\begin{align}
&\ddot{\phi} + 3 H \dot{\phi} - \frac{\nabla^2 \phi}{a^2}+\frac{\partial V}{\partial \phi} + \frac{\beta}{4 f_a}
F_{\mu \nu} \widetilde{F}^{\mu \nu} = 0, \label{eq:phi}\\
&\ddot{\bf A} + H \dot{\bf A} - \frac{\nabla^2 {\bf A}}{a^2}+m_{\gamma'}^2 {\bf A} 
- \frac{\beta}{f_a a} \left(\dot{\phi} \nabla \times {\bf A}  \right. \nonumber \\ 
&~\left. - \nabla \phi \times \left(\dot{\bf A} - \nabla A_0\right)\right)=0,\label{eq:A}
\end{align}
where the overdot is the derivative with respect to time $t$, 
$a(t)$ the scale factor, $H$ the Hubble parameter, and $\nabla^2 = \partial_i^2$. 
The evolution of $A_0$ is determined by the Lorenz gauge condition, $\partial_\mu(\sqrt{-g} A^\mu ) = 0$,
which directly follows from the equation of motion.

Suppose that the spatially homogeneous axion starts to oscillate with an initial
amplitude $\phi_i$ when $3H \sim m_a$ in the radiation-dominated era. 
Then the equation of motion of ${\bf A}$ is reduced to 
\begin{align}
&\ddot{\bf A}_{\bf k, \pm} + H \dot{\bf A}_{\bf k, \pm}
+\left(m_{\gamma '}^2 + \frac{k^2}{a^2} \mp \frac{k}{a} \frac{\beta \dot{\phi}}{f_a}\right) {\bf A}_{\bf k, \pm} = 0
\end{align}
in Fourier space, where $k = |{\bf k}|$ denotes the comoving wave number,
and the subscript $\pm$ indicates the helicity of the transverse mode. 
One can see that one of the helicity components 
with $k/a \sim \beta |\dot{\phi}|/2 f_a$ becomes tachyonic if $m_{\gamma'} < \beta |\dot{\phi}|/2 f_a$,
and such dark photons are efficiently produced by the tachyonic instability soon after the axion starts to oscillate. 
Note that only the transverse mode of the dark photon is coupled to 
the spatially homogeneous axion. 
After the energy density of dark photons becomes comparable to the axion, the system enters a non-linear regime. The energy stored in the axion zero mode transfers to both transverse and longitudinal components of dark photons as well as the axion non-zero mode. As we shall see shortly, however, the dark photon production effectively stops soon after the system enters the non-linear regime. The dark photon (physical) momentum has a characteristic peak at $\sim 10^{-2} \beta m_a$ at that moment, where the numerical prefactor also depends on $\beta$ but we confirmed its validity for $\beta$ between $35$ and $50$.
The dark photon abundance is related to the initial axion abundances as
\begin{align}
\frac{\rho_{\gamma'}}{s} &\simeq  \frac{m_{\gamma'}}{10^{-2} \beta m_a} \left.\frac{\rho_a}{s}\right|_{3H = m_a},
\end{align}
where $s$ is the entropy density, and $\rho_{\gamma'}$ and $\rho_a$ 
the energy densities of dark photon and axion, respectively.
Here we have approximated 
that most of the initial axion energy transfers to dark photons that are relativistic at the production. 
In terms of the density parameter,
it is given by
\begin{align}
\Omega_{\gamma '}h^2 &\simeq 0.2\, \theta^2 
\left(\frac{40}{\beta}\right)
\left(\frac{m_{\gamma'}}{10^{-9}{\rm eV}}\right)
\left(\frac{10^{-8} {\rm eV}}{m_{a}}\right)^{\frac{1}{2}}
\left(\frac{f_a}{10^{14}{\rm GeV}}\right)^2,
\label{eq:Omega}
\end{align}
where we set the relativistic degrees of freedom $g_*(T) = 60$. 
As we shall see in the next section, the production efficiency depends on the dark photon mass,
and some amount of the axions always contribute to DM. 

Note that the axion acquires its quantum fluctuations during inflation, 
which induce isocurvature perturbation of dark photons after the tachyonic production. 
The amplitude of the isocurvature perturbation is given by ${\cal P}_S^{1/2} = H_I/(\pi f_a \theta)$. Using $f_a \theta$ in Eq.\,(\ref{eq:Omega}) with $\Omega_{\gamma'}=\Omega_{\rm DM}$, the current constraint on the isocurvature perturbation~\cite{Akrami:2018odb} limits the inflation scale as
\begin{align}
H_I < 2 \times 10^9\,{\rm GeV} \left(\frac{\beta}{40}\right)^{\frac{1}{2}} \left(\frac{m_{\gamma'}}{10^{-9}\,{\rm eV}}\right)^{-\frac{1}{2}} \left(\frac{m_a}{10^{-8}\,{\rm eV}}\right)^{\frac{1}{4}}.
\end{align}
Thus, relatively low-scale inflation models are required. In this sense, our scenario is complementary to the production mechanism using inflationary fluctuations \cite{Graham:2015rva} which typically requires higher $H_I$.

\begin{figure*}[tp]
\centering
\includegraphics[width=0.45\textwidth]{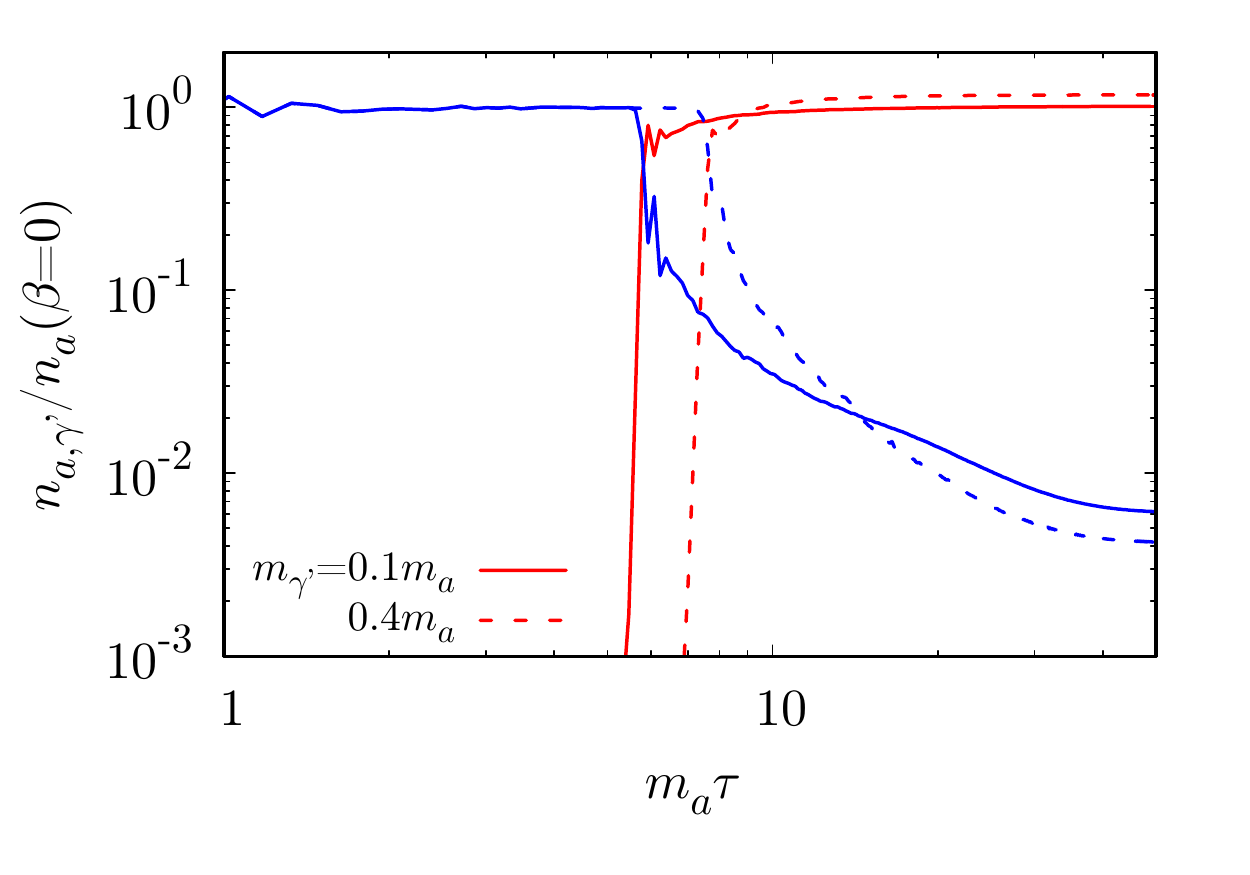}
\qquad
\includegraphics[width=0.45\textwidth]{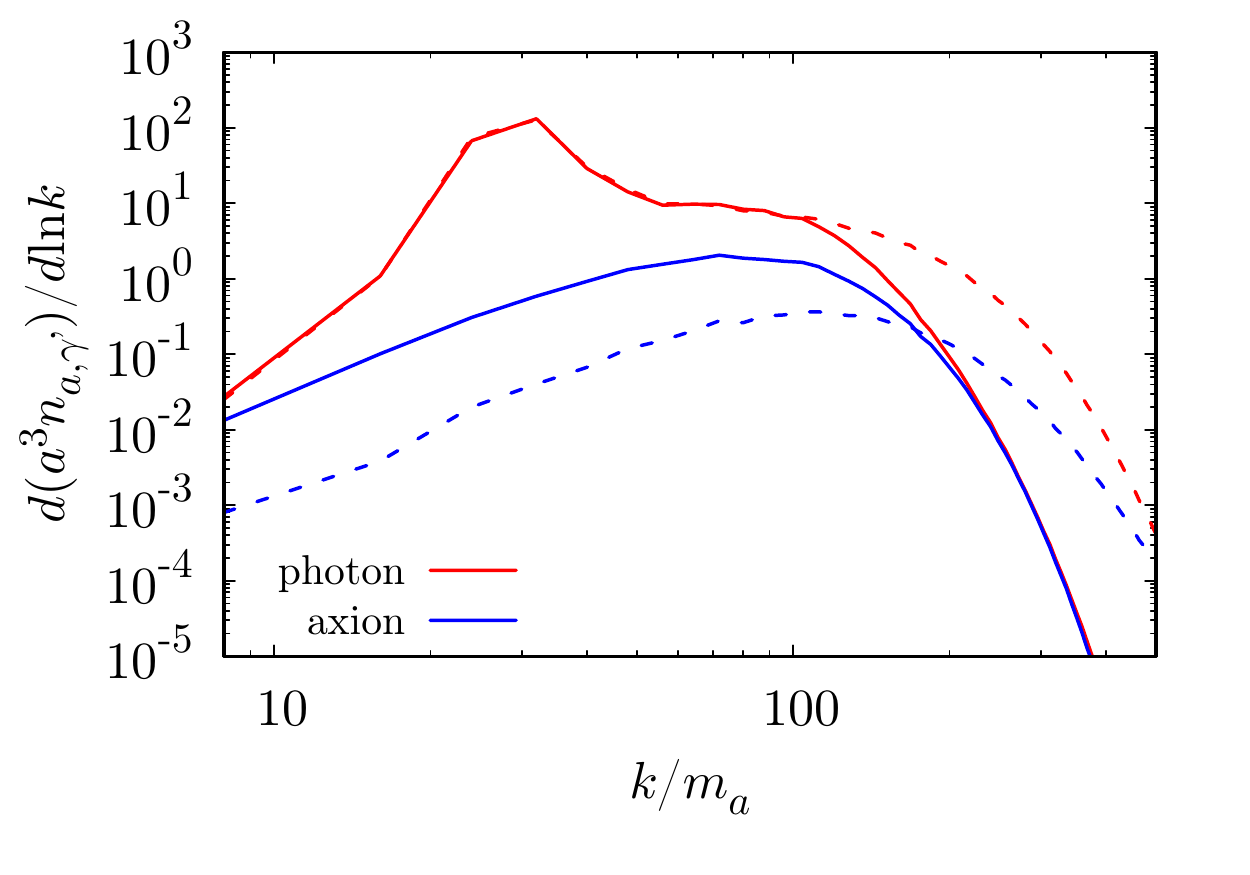}
\caption{
  Left: 
  The evolution of the number densities of axion (blue) and dark
  photon (red) normalized by that of the axion without tachyonic
  production. We show two benchmarks, $m_{\gamma'} = 0.1 m_a$ (solid
  line) and $0.4m_a$ (dashed line).
  Right:
  The power spectra of the number density of axion (blue) and dark
  photon (red) with $\mdp=0.1m_a$. We show two snapshots at $m_a \tau
  = 10$ (solid) and
  50 (dashed).
  Other parameters were taken to be 
  $\beta=40$, $f_a = 10^{14}$ GeV, $m_a = 10^{-8}$ eV for both plots. 
}
\label{fig:evolve}
\label{fig:spectrum}
\end{figure*}

Let us here briefly comment on an additional condition for the above
production mechanism to work. 
After the tachyonic production ends, the
dark photon field amplitude is as large as $f_a$. More generally,
light dark photon DM, when extrapolated to the early
universe, would have had large field values.
Consequently, even tiny shift-symmetry violating couplings can have
a dramatic effect on the dynamics.
If the mass of the dark photon arises from a Higgs mechanism, the
dark photon field also backreacts on the Higgs potential.
Even without the Higgs mechanism, a quartic self-coupling is allowed
for a St\"uckelberg dark photon. 
A generic expectation for the dark photon quartic in either case is
$\sim g_D^4$.
So far we have ignored any such couplings of the dark photon. Any UV
completion of our scenario must explain why these couplings are small.
We further discuss this issue in the supplementary material.



\begin{figure}[tp]
\centering
\includegraphics [width = 8cm, clip]{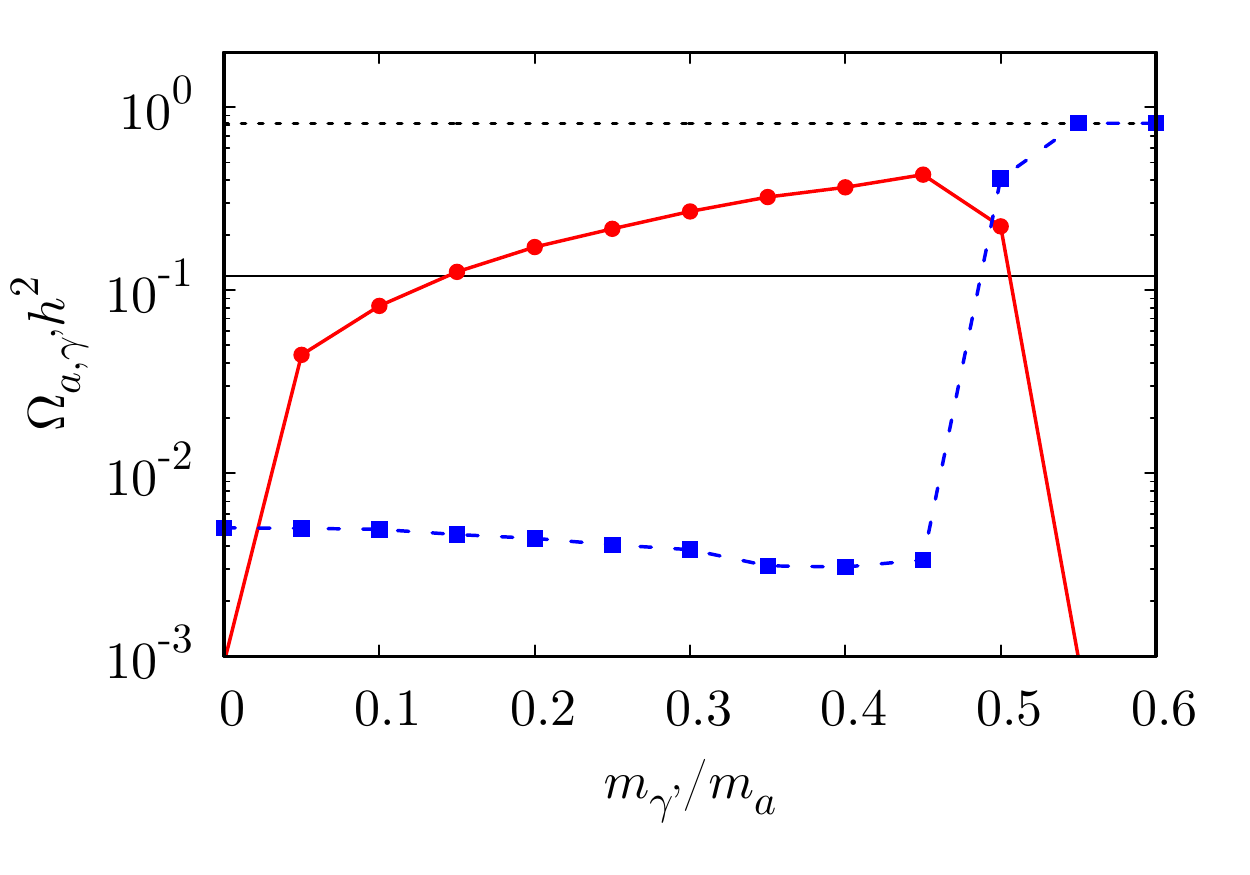}
\caption{
The relic density parameter of axion (blue square) and dark photon (red circle) as a function of $m_{\gamma'}/m_a$.
 The horizontal solid and dotted lines represent the observed DM density parameter and the axion density parameter  without tachyonic production.
 The values of $\{\beta$, $f_a$ $m_a\}$ are as
 in~\cref{fig:evolve}.
}
\label{fig:Omega}
\end{figure}
\section{Numerical results}
We have solved directly the equations of motion (\ref{eq:phi}) and
(\ref{eq:A}) by performing lattice numerical simulations on a cubic
lattice with periodic boundary conditions with $128^3$ points and
initial comoving lattice spacing $ = (\pi/512)m_a^{-1}$.  As reference values we have
taken $\beta=40$, $m_a=10^{-8} {\rm eV}$, and $f_a = 10^{14}$\,GeV,
and adopt the initial condition $\phi_i=f_a$  at the conformal time
$\tau_i = 0.1m_a^{-1}$. The scale factor is normalized as $a(\tau)
=\tau/\tau_i$, and the Hubble parameter is given by $H = \tau_i/\tau^2$ in the radiation dominated era.
 We adopt initial
fluctuations of the dark photon given by quantum vacuum fluctuations
following the Rayleigh distribution in Fourier space with the
root-mean-square amplitude
\begin{equation}
\sqrt{\langle |A_{\bf k}|^2 \rangle} = \frac{1}{\sqrt{2\omega_k}}~~\text{with}~~ \omega_k = \sqrt{(k/a)^2+m_{\gamma'}^2}.
\end{equation}
The initial value at each spatial point is obtained by the inverse-Fourier transformation.

We show in \cref{fig:evolve} (left) the time evolution of the axion
(blue) and dark photon (red) number densities normalized by that of
the axion in the case of no dark photon production. One can see that
the axion number density abruptly drops when the dark photon becomes
comparable to the axion in number at $m_a\tau \simeq 6 - 8$, and that
the final dark photon number density is more than $10^2$ times larger
than that of the axion. Most of the initial axion energy
is efficiently transferred to dark photons.

In \cref{fig:spectrum} (right) we show the power spectra of the comoving
number densities of the axion (blue) and the dark photon (red) at the
conformal time $m_a \tau = 10$ (solid) and 50 (dashed).  The dark
photon spectrum has a prominent peak corresponding to the fastest
growing mode of the tachyonic instability,  and the peak continues to
persist after the tachyonic growth is saturated at $m_a \tau = 6$.
The physical momentum of the peak at this time is $k_{\rm phys}
\sim 0.5 m_a$.
The backreaction is not efficient for
dark photons near the peak, and their production effectively stops
soon after $m_a \tau = 6$. 
On the other hand, dark photons with higher momentum modes with
$k_{\rm phys} > m_a$ ($k/m_a > 100$ at $m_a \tau = 10$) continue to
interact and axions with lower momentum are converted to
dark photons and axions with higher momenta. This, however, does not
affect the final dark photon abundance.

\begin{figure*}[tp]
  \centering
  \includegraphics[width=0.45\textwidth]{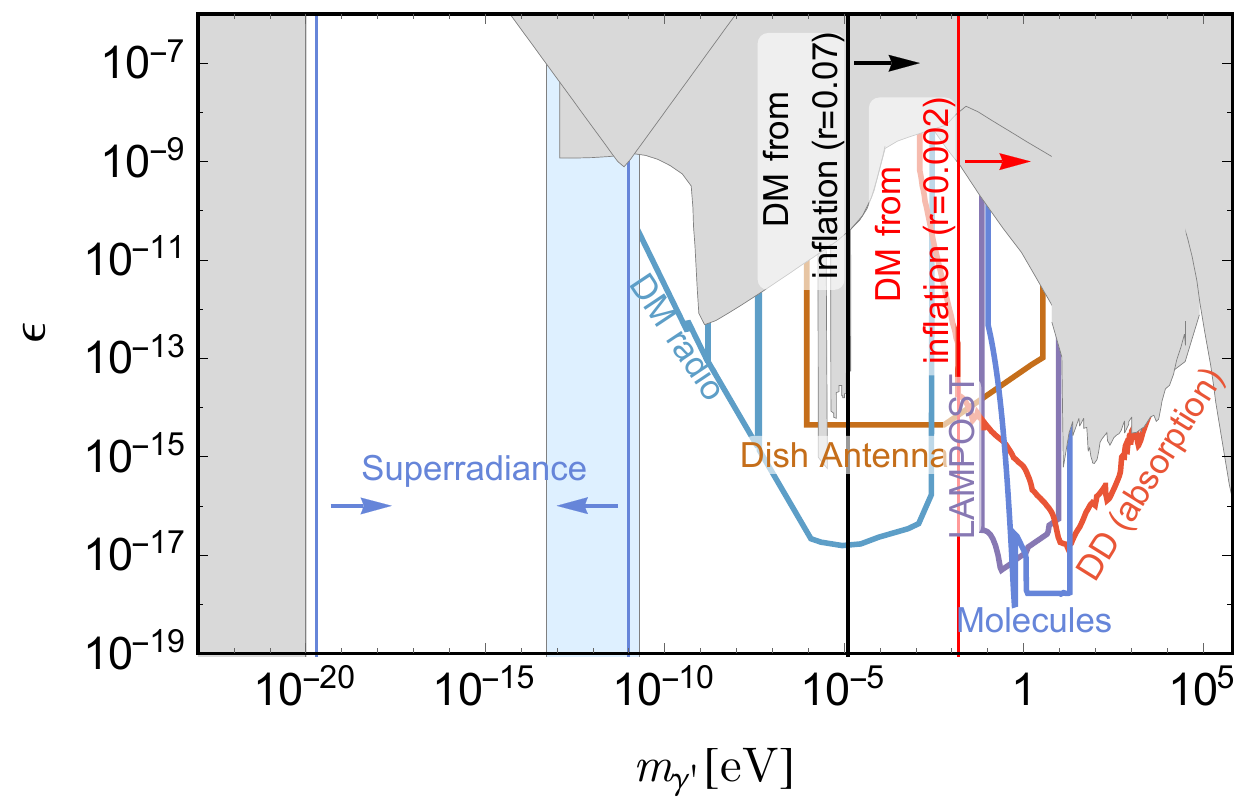}
  \quad
  \includegraphics[width=0.45\textwidth]{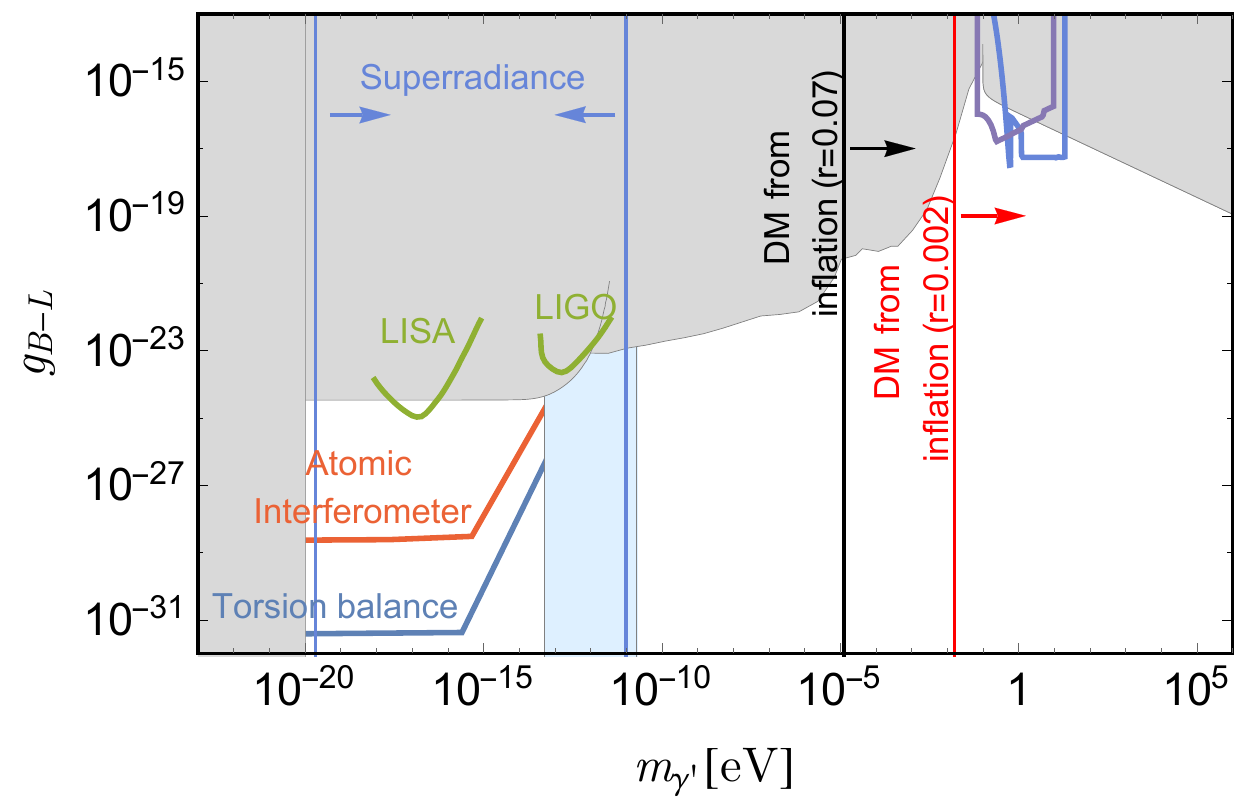}
  \caption{ Current constraints (gray) and future prospects for
  massive spin-1 particle. Left: coupling to the SM through kinetic
mixing.
Right: a weakly coupled $B-L$ gauge boson.}
  \label{fig:pheno}
\end{figure*}

In \cref{fig:Omega} we show the relic density parameters of axion (blue
square) and dark photon (red circle) as a function of $m_{\gamma'}$.
One can see that the dark photon abundance linearly increases  up to
$m_{\gamma'} = 0.45m_a$ and sharply drops for heavier $m_{\gamma'}$. 
This is because the tachyonic growth no longer takes place for the mass heavier than the peak momentum,
 $m_{\gamma'} \gtrsim 10^{-2} \beta m_a \sim 0.5 m_a$.
 
We have also performed lattice simulations with different sets of the
parameters and found similar behavior if the tachyonic production is
efficient enough for the system to enter the nonlinear regime. 
If we take a closer look, however, the peak of the dark photon spectrum
becomes less prominent as $\beta$ increases.
On the other hand, for $\beta \lesssim 30$,
the tachyonic production is so inefficient that the dark photon number
remains smaller than axions.

\section{Phenomenology}

The mechanism presented in this letter does not rely on any couplings
to the SM. Thus, the only constraints that apply to the mechanism
generally are superradiance~\cite{Arvanitaki:2009fg,Arvanitaki:2010sy,Baryakhtar:2017ngi,Cardoso:2017kgn,Cardoso:2018tly} and structure formation~\cite{Hui:2016ltb,Irsic:2017yje}.
The superradiance constraints potentially apply to both the dark photon
directly, as well as to the axion field. 
We focus here on the direct constraints on dark photons. 
We have shown these constraints in light blue
in~\cref{fig:pheno}.
Note that superradiance constraints require the self-interactions of
the gauge boson to be small, which we discuss further in the
supplementary material. The Lyman-$\alpha$ constraint on fuzzy DM~\cite{Irsic:2017yje} 
directly applies to dark photons.
Considering that dark photons are mildly relativistic at the production, we adopt $\mdp \gtrsim
10^{-20}$\,eV as the lower bound.

More generally, the dark
photon could have additional couplings to the SM, giving rise to a
rich phenomenology at current and future experiments~\cite{Pospelov:2008jk,Redondo:2008ec,Arias:2012az,Pospelov:2018kdh}.
We focus on two specific examples: that of a dark photon kinetically
mixed with the hypercharge gauge boson, and a $B-L$ gauge boson with a
very small gauge coupling.

There are a number of ongoing and future experiments that look for
kinetically mixed dark photons over a large range of their masses
(see~\cite{Essig:2013lka} for a review).  Current constraints come
from Lyman-$\alpha$ measurements~\cite{Irsic:2017yje}, black hole
superradiance~\cite{Baryakhtar:2017ngi,Cardoso:2017kgn,Cardoso:2018tly},
galactic heating~\cite{Dubovsky:2015cca}, late resonant conversions of
photons to dark photons, CMB distortions,
$N_{\rm eff}$~\cite{Arias:2012az}, direct
detection~\cite{An:2013yua,An:2014twa,Aguilar-Arevalo:2016zop}, stellar
cooling~\cite{An:2013yfc,Redondo:2013lna,Hardy:2016kme}, and decays of
dark photons to X-rays. There are additional constraints in a narrower
mass range from ADMX~\cite{Wagner:2010mi} and from  a dish antenna
experiment~\cite{Horns:2012jf,Knirck:2018ojz}. Other astrophysical
constraints such as those from astronomical
ephemerides~\cite{Fukuda:2018omk} can also be translated to the dark
photon case.

New experiments seek to cover new parameter space for the dark photon.
These include
DM Radio~\cite{Chaudhuri:2014dla,Silva-Feaver:2016qhh},
resonant absorption in molecules~\cite{Arvanitaki:2017nhi},
LAMPOST~\cite{Baryakhtar:2018doz},
dish antenna experiments~\cite{Horns:2012jf,Dobrich:2015tpa},
direct detection~\cite{Hochberg:2016ajh,Hochberg:2016sqx,Bloch:2016sjj}, and 
a resonant microwave cavity search~\cite{Parker:2013fxa}.
In~\cref{fig:pheno} we show constraints and prospects for a dark
photon in the $m$-$\epsilon$ plane. 
We see that there are regions of
parameter space $\mdp  \sim 10^{-9} \eV, \, \epsilon\sim 10^{-12}$ 
where the DM radio would be sensitive, but cannot rely on inflationary
perturbations to populate DM. Thus, this is a prime target
space for our mechanism.

If the dark photon couples to a different current than the electromagnetic
current (which we take to be
$B-L$ for concreteness), then matter in the universe may not be charge
neutral under this new force. This leads to additional interesting
constraints from tests of equivalence principle
violation and fifth
forces~\cite{Talmadge:1988qz,Smith:1999cr,Williams:2004qba,Wagner:2012ui,Berge:2017ovy,Fayet:2017pdp,Touboul:2017grn}.
There are also additional constraints from decays of the $B-L$
gauge boson into neutrinos; we conservatively constrain the lifetime
to be more than the age of the universe, but a tighter bound can
likely be derived~\cite{Enqvist:2015ara}. 
These forces can also be tested in future experiments such as atomic
interferometers~\cite{Graham:2015ifn,Kalaydzhyan:2018zsx}, new torsion pendulum
experiments, or with LIGO and
LISA~\cite{Graham:2015ifn,Pierce:2018xmy}. This gives another target
for our mechanism, $\mdp\sim$ $10^{-20}$--$10^{-15}\eV$, with a very
small gauge coupling $g_{B-L} < 10^{-25}$. 

We emphasize that these benchmark models are chosen for concreteness,
and our mechanism may be extended to a wide class of dark photon
models.

\section{Conclusion}
The nature of DM remains a pressing mystery in fundamental
physics. An important possibility is that DM is made up of
very light, spin-1 particles which form a coherent, classical field. 
In this letter we provide a mechanism for the generation of the
abundance of such DM candidates. Our mechanism works over the
entire range of DM masses which are not otherwise
constrained, in contrast with existing mechanisms which only work for
a range of masses, or for a very fine-tuned choice of couplings. Thus,
our mechanism puts the motivation for dark photon DM on a
firmer footing. 

\medskip

\paragraph*{\bf Note added:} As this paper was being completed we became aware of overlapping work in preparation from other groups \cite{Co:2018, Dror:2018, BasteroGil:2018}.

\bigskip

\paragraph*{\bf Acknowledgments:}
We thank N.~Craig for a useful comment.
FT thanks H.~Yoshino for useful communications on superradiance.
PA and MR thank Galileo Galilei Institute for Theoretical Physics for
their hospitality and the INFN for partial support during the completion of this work.
PA would also like to
thank the Kavli Institute for Theoretical Physics for
hospitality, supported in part by the National Science Foundation under Grant No. NSF PHY17-48958.
FT thanks
the hospitality of MIT Center for Theoretical Physics and Tufts
Institute of Cosmology where this work was done. 
The work of PA is supported by the NSF grants
PHY-0855591 and PHY-1216270. MR is supported in part by the NASA ATP Grant NNX16AI12G and by the DOE Grant DE-SC0013607. 
TS is supported by JSPS KAKENHI Grant Number JP15H02082, JP18H04339 and JP18K03640.
FT is supported by JSPS KAKENHI Grant Numbers JP15H05889, 
JP15K21733, JP17H02878, JP17H02875, Leading Young Researcher Overseas
Visit Program at Tohoku University, and WPI Initiative, MEXT, Japan.
NK acknowledges the support by Grant-in-Aid for JSPS Fellows.
This work was supported by a grant from the Simons Foundation (341344, LA).

\bibliography{ref}

\clearpage
\newpage
\maketitle
\onecolumngrid
\begin{center}
\textbf{\large Relic Abundance of Dark Photon Dark Matter} \\ 
\vspace{0.05in}
{ \it \large Supplementary Material}\\ 
\vspace{0.05in}
{Prateek Agrawal, Naoya Kitajima,  Matthew Reece, Toyokazu Sekiguchi, and Fuminobu Takahashi}
{ }

\end{center}
\counterwithin{figure}{section}
\setcounter{equation}{0}
\setcounter{figure}{0}
\setcounter{table}{0}
\setcounter{section}{0}
\renewcommand{\theequation}{S\arabic{equation}}
\renewcommand{\thefigure}{S\arabic{figure}}
\renewcommand{\thetable}{S\arabic{table}}
\newcommand\ptwiddle[1]{\mathord{\mathop{#1}\limits^{\scriptscriptstyle(\sim)}}}

\section{Models}
\label{sec:models}
A number of model building avenues for obtaining dark photons have
been
pursued~\cite{Abel:2008ai,Goodsell:2009xc,Feng:2014eja,Feng:2014cla,Harigaya:2016rwr}.
The key model-building questions for our purposes are the origin of
the dark photon mass, the absence of quartic couplings for the dark
photon, 
the origin and size of the dark photon coupling
to the axion and a possible coupling of the dark photon to the
SM.

In effective field theory, a very light dark photon could either arise
from the Higgs mechanism (including its generalized form, dynamical
symmetry breaking) or from the St\"uckelberg mechanism. However, in
string theory one finds that very light St\"uckelberg masses are
always correlated with a low string scale, and imposing a lower bound
of $\sim {\rm TeV}$ on the string scale implies that dark photon
masses below the meV scale must arise from the Higgs mechanism~\cite{Goodsell:2009xc, Reece:2018zvv}. 

\subsection{Quartic couplings: expected size and bound on gauge
coupling}

The theory of a massive spin-1 boson lacks gauge invariance, and so in
general allows a variety of new interaction terms, such as
\begin{align}
  \mathcal{L}
  &\supset
\lambda_{\gamma'} (A_\mu A^\mu)^2
\,. 
  \label{eq:quartic-definition}
\end{align}
Such terms are dangerous for our
mechanism, because they make it energetically costly to populate a
large occupation number for the dark photon, eventually backreacting
and shutting off the mechanism. Thus the success of our mechanism
depends on having a small quartic coupling. In fact, it requires not
only a small $\lambda_{\gamma'}$, but also small gauge-invariant
interactions. 

The non-gauge-invariant $\lambda_{\gamma'}$ interaction must vanish in
the limit $\mdp \to 0$, so we expect that $\lambda$ is suppressed by
powers of $\mdp$. A generic expectation is that in the $g_D \to 0$
limit, the longitudinal mode of the photon becomes a compact scalar
$\theta$ with a Lagrangian of the form \begin{equation} f_\theta^2
  (\partial_\mu \theta)^2 + c (\partial_\mu \theta)^4 + \ldots,
\end{equation} with $c \sim 1$ and $\mdp = g_D f_\theta$. We couple this
theory to the gauge field with the replacement $\partial_\mu \theta
\to \partial_\mu \theta + g_D A_\mu$, which makes it apparent that we
expect the coefficient to be of order 
\begin{equation}
  \lambda_{\gamma';{\rm generic}} \sim g_D^4 \sim
  \left(\frac{\mdp}{f_\theta}\right)^4.  
\end{equation}
This shows
that, as expected, $\lambda \to 0$ when $\mdp \to 0$, but at {\em
fixed} $f_\theta$. On the other hand, if $g_D$ is not very small the
quartic is not necessarily very suppressed. In the concrete case of
the abelian Higgs model with a Higgs of charge $q_h$, VEV $v_h$, and
quartic coupling $\lambda_h$, the quartic is induced by Higgs
exchange: 
\begin{align} 
  \lambda_{\gamma';{\rm Higgs}} 
  &= \frac{(g_D^2
  q_h^2 v_h)^2}{2 m_h^2} = \frac{g_D^4 q_h^4}{4 \lambda_h}.
  \label{eq:quarticfromhiggs} 
\end{align} 
As in the generic case,
the quartic scales with four powers of the small coupling $g_D$.

The second type of self-interaction that is generically present is a
gauge-invariant Euler-Heisenberg type interaction, $(F^2)^2$ or $(F
{\widetilde F})^2$. We can examine the size of such a term generated
by integrating out fermions of mass $m_C$ and charge $q_C$, and then
interpret this as an effective simple quartic $\lambda_{\rm eff}$ by
taking the derivatives to be of order $\mdp$: 
\begin{equation}
  \frac{1}{4!} \frac{g_D^4 q_C^4}{16\pi^2m_C^4}
  \left(\frac{4}{15}(F_{\mu \nu}F^{\mu \nu})^2 + \frac{7}{15}(F_{\mu
  \nu}{\widetilde F}^{\mu \nu})^2\right) \quad \Rightarrow \quad
  \lambda_{\rm eff} \sim \frac{g_D^4 q_C^4}{16\pi^2}
  \left(\frac{\mdp}{m_C}\right)^4.  
\end{equation} 
As with the simple
quartic case, $\lambda_{\rm eff}$ is suppressed by $g_D^4$; however, it
is additionally suppressed by small couplings and by the unknown
masses of heavy charged particles. Hence, it is reasonable to expect
that for a typical dark photon model, simple $A_\mu^4$-type quartic
couplings dominate over the Euler-Heisenberg coupling.

If we wish for the quartic coupling $\lambda_{\gamma'}$ to play no
role in the dynamical mechanism we discuss, it must obey a stringent
inequality. First, we would like the energy density stored in the
quartic term to be unimportant compared to that in the mass or kinetic
terms, which gives us
\begin{equation}
\lambda_{\gamma'} \lesssim \frac{m_{\gamma'}^2}{\langle A_\mu^2 \rangle} \lesssim \frac{m_{\gamma'}^4}{m_a^2 f_a^2}.    
     \label{eq:smallquartic}
\end{equation}
Second, if we assume that the dark photon mass arises from the Higgs
mechanism, we would like the photon field to not significantly
backreact on the Higgs mass, i.e.
\begin{align}
g_D^2 q_h^2 \langle A_\mu^2 \rangle \ll \lambda_h v_h^2,
\end{align}
which using \eqref{eq:quarticfromhiggs} leads to the same inequality
\eqref{eq:smallquartic}. Because the size of the Higgs quartic
$\lambda_h$ is constrained by unitarity, the only way that
\eqref{eq:smallquartic} can be obeyed in a theory in which the dark
photon mass arises from a Higgs mechanism is to have an exceptionally
small gauge coupling: the small photon mass arises from a relatively
large VEV for the dark Higgs multiplied by a correspondingly small
coupling: 
\begin{align} 
  v_h 
  &\gtrsim
  1\GeV\,
  \left(\frac{1}{\lambda_h}\right)^{1/4}
  \left(\frac{m_a}{\mueV}\right)^{1/2}
  \left(\frac{f_a}{10^{14}\GeV}\right)^{1/2}
  \\
  g_D 
  &\lesssim
  10^{-15}\,
  \left(\frac{\lambda_h}{q_h^4}\right)^{1/4}
  \left(\frac{\mdp}{\mueV}\right)
  \left(\frac{\mueV}{m_a}\right)^{1/2}
  \left(\frac{10^{14}\GeV}{f_a}\right)^{1/2}
  .
  \label{eq:gaugecouplingbound} 
\end{align}
These are awkward numbers to explain in a fundamental
theory, but the coupling is not so small as to be excluded by Weak
Gravity Conjecture arguments, for instance.

\subsection{Coupling of axions to dark photons}

A further challenge arises when we attempt to explain the size of the
axion-dark photon coupling that we rely on for particle production,
\begin{align}
  \mathcal{L}
  &=
  \beta \frac{\phi}{f_a} F_D \widetilde{F}_D
\end{align}
with $\beta \sim \mathcal{O}(10$--$100)$ where $f_a$ is the field range of
$\phi$. For a theory of a compact $U(1)$ gauge field and a single
compact axion, the coefficient $\beta$ is actually quantized. A useful
parametrization is
\begin{equation}
\beta = k j \frac{\alpha_D}{8 \pi},
\end{equation}
highlighting the coupling's dependence on the gauge coupling and two
potential sources of
enhancement. Here $k$ denotes the anomaly integer that may be large
due to the presence of a fermion with a large charge or due to a large
number of flavors. Perturbativity of the gauge theory requires that
$k\alpha_D \lesssim 1$, so $k$ by itself cannot yield $\beta > 1$.  As
is familiar from other contexts where this coupling
arises~\cite{Agrawal:2018mkd}, to get $\beta \sim 100$, we need to
invoke a form of alignment \cite{Kim:2004rp} or its iterated version, clockwork \cite{Choi:2014rja,Choi:2015fiu,Kaplan:2015fuy}.  Here $j$ denotes an
enhancement of the field range over the fundamental period of the
axion using alignment/monodromy. 

The requirement of not generating a large quartic coupling from the
Higgs mechanism compounds this problem: because $\alpha_D \lesssim
10^{-30}$, the enhancement factors $j$ and $k$ must be enormous.
Furthermore, if we generate the coupling by integrating out charged
fermions that also carry PQ charge, then they can run in loops and
generate Euler-Heisenberg-type quartic couplings, which may be so
large as to again pose a dynamical problem for our mechanism.  Let us
try to estimate whether viable parameter space exists. If we have a
set of PQ-charged fermions with dark charge $q_\Psi$ obtaining a mass
from PQ breaking, we expect to generate the axion-dark photon coupling
with
\begin{equation}
k \sim \sum q_\Psi^2.
\end{equation}
We can then enhance this by a factor of $j$ in a fundamental theory of multiple axions. An example model is,
\begin{align}
  \mathcal{L}
  &=
  \frac12(\partial a)^2
  + \frac12(\partial b)^2
  + \frac{ k \alpha_D }{8\pi F} a F_D \widetilde{F}_D
  + \mu^4 \cos\left(\frac{b}{F}\right)
  + \Lambda^4 \cos\left(\frac{a + j b}{F}\right)
  \,.
\end{align}
where $\Lambda \gg \mu$.
Identifying the field range of the light
field $f = F \sqrt{j^2+1}$,
\begin{align}
  \mathcal{L}_{\rm eff}
  &=
  \frac12(\partial \phi)^2
  + \frac{ k j \alpha_D }{8\pi f} \phi F_D \widetilde{F}_D
  + \mu^4 \cos\left(\frac{\phi}{f}\right)
  .
\end{align}
In this way, we can arrange for the masses of the PQ-charged fermions
to lie at the scale $F \sim f/j$. Then the corresponding
Euler-Heisenberg term will scale as
\begin{align}
\frac{\alpha_D^2}{(4!) F^4} 
\sum q_\Psi^4 (F_{\mu \nu}^2)^2 
&\sim 
\frac{\alpha_D^2 k^2 j^4}{(4!) f_a^4} 
(F_{\mu \nu}^2)^2 \sim \frac{(8 \pi \beta j)^2}{(4!) f_a^4} 
(F_{\mu \nu}^2)^2.    
\label{eq:EHcoupling}
\end{align}
Perturbativity of the gauge theory requires that $j \gtrsim \beta$. We
can estimate an effective quartic interaction $\lambda_{\gamma'}$ by
taking $F \sim p A$ where $p \lesssim \beta m_a$ is the characteristic
momentum involved in our dark photon production mechanism. In that
case, applying the constraint \eqref{eq:smallquartic} gives
\begin{equation}
\frac{(8\pi \beta j)^2}{4!} \left(\frac{\beta m_a}{f_a}\right)^4 \lesssim \frac{m_{\gamma'}^4}{m_a^2 f_a^2}.
\end{equation}
Let us consider two different limits to assess the importance of this
bound. In the first case, we will take the minimum amount of
clockwork: when $\alpha_D k \sim 1$ to saturate the perturbative
unitarity constraint on the gauge theory, we must take $j \sim \beta$,
and the inequality becomes:
\begin{equation}
\beta \lesssim 10^7 \left(\frac{f_a}{10^{14}~{\rm GeV}}\right)^{1/4} \left(\frac{\mdp}{\mu{\rm eV}}\right)^{1/2} \left(\frac{\mu{\rm eV}}{m_a}\right)^{3/4}.
\end{equation}
We have tested our mechanism at much smaller values of $\beta$, so
this is not a significant constraint. However, note that this is the
case where we have made minimal use of the clockwork mechanism, taking
$j \sim \beta$, in which case we require enormous $k$.

Now consider the opposite limit, taking $k$ as small and $j$ as large as possible. As argued in \cite{Agrawal:2018mkd}, the clockwork mechanism leads to the constraint that the scale of the axion potential is bounded by the fundamental decay constant $f_a / j$, 
\begin{equation}
j \lesssim \sqrt\frac{f_a}{m_a}.
\end{equation}
Otherwise, axion scattering will violate perturbative unitarity. Suppose that we saturate this bound. Then we find that the Euler-Heisenberg interaction \eqref{eq:EHcoupling} is safe from the quartic constraint \eqref{eq:smallquartic} provided
\begin{equation}
\beta \lesssim 4 \times 10^4 \left(\frac{f_a}{10^{14}~{\rm GeV}}\right)^{1/6} \left(\frac{\mdp}{\mu{\rm eV}}\right)^{2/3} \left(\frac{\mu{\rm eV}}{m_a}\right)^{5/6}.
\end{equation}
Thus we can exploit the clockwork mechanism to the maximum extent
compatible with unitarity while still remaining safe from the quartic
constraint. However, if $g_D \sim 10^{-15}$ and $j \sim \sqrt{f_a/m_a}
\sim 10^{15}$, we still require an unexplained large number $k \sim
10^{15}$. This is, as far as we know, logically consistent, but it
suggests the need for a better model.

\subsection{Coupling to the SM}
The phenomenology of the dark photon depends sensitively on its
couplings to the SM. We consider two benchmark models for dark photon
couplings---a kinetic mixing with the hypercharge boson, and coupling
to the $B-L$ current.

Kinetic mixing is a well-studied mechanism for the dark photon to talk with the SM. In the simplest models, we expect the size of the
kinetic mixing parameter to be,
\begin{align}
  \epsilon
  &\simeq
  \frac{e g_D}{16\pi^2}
  \lesssim
  10^{-18}\,
  \left(\frac{\lambda_h}{q_h^4}\right)^{1/4}
  \left(\frac{\mdp}{\mueV}\right)
  \left(\frac{\mueV}{m_a}\right)^{1/2}
  \left(\frac{10^{14}\GeV}{f_a}\right)^{1/2}
  \,.
\end{align}
This level of
kinetic mixing is a little too small to be experimentally accessible
(\cref{fig:pheno}) except perhaps towards higher dark photon masses.
However, as we note above that the model includes
fermions with large charges in order to obtain the requisite coupling
of the axion with dark photons. The same fermions can induce a much
larger kinetic mixing. For example, if there is a fermion with charge
$(1, \sqrt{k})$ under $(U(1)_Y, U(1)_D)$, then we obtain the estimate,
\begin{align}
  \epsilon
  &\simeq
  \frac{\sqrt{k} e g_D}{16\pi^2}
  \sim
  10^{-10}
\end{align}
where in the last simequality we have used the estimate for $k$ and
$g_D$ obtained above for
$\mueV$ dark photons. Therefore,
within the class of models with large charges it is possible to get
kinetic mixing parameters in all of the experimentally interesting
region.

Another well-motivated possibility is that the DM is made up
of $B-L$ gauge bosons.  
Due to very stringent constraints on new long range forces for SM
fields, its gauge coupling is constrained to be extremely small. This
meshes well with our bound derived from the quartic coupling;
the constraints on the gauge coupling are typically stronger than
those derived from the restrictions on the quartic coupling. 
As noted above, the coupling to the axion still requires a very large
charged field in the theory.

\subsection{Generating large charges through Abelian Clockwork}
From above we see that we require very large relative integer charges
for our mechanism.
An attractive set-up to generate such a large charge is the clockwork
mechanism, as applied
to $U(1)$ gauge symmetries.
We imagine a string of $U(1)$ gauge symmetries, which are all broken
down to a diagonal $U(1)$ by VEVs of $N-1$ charged
scalars, with $\phi_{(i,i+1)}$ charged as $(1,q)$ under $(U(1)_i,
U(1)_{i+1})$.
\begin{align}
  \mathcal{L}
  &=
   |(\partial_\mu - i A_N^\mu) h|^2 - V(h)
  + J_\mu A_1^\mu
 +\sum_i
 -\frac{1}{4g_i^2} F_i F_i
  + |D_\mu \phi_{(i,i+1)}|^2 - V(\phi_{(i,i+1)})
  \,.
\end{align}
Here $h$ is the Higgs that provides a small mass to the dark
photon that would be otherwise massless, 
and $J_\mu$ is a current that we want to couple to the dark
photon with an enhanced strength.
The effective Lagrangian is,

\begin{align}
  \mathcal{L}_{\rm eff}
  &=
 -\frac{1}{4g_D^2} F_D F_D
   + |(\partial^\mu - i A_D^\mu) h|^2 - V(h)
  + q^N J_\mu A_D^\mu
  \,.
\end{align}

The dark photon gauge coupling is given as,
\begin{align}
  \frac{1}{g_D^2}
  &=
  \sum
  \frac{1}{g_i^2}
  {q^{2i}}
  =
  \frac{1}{g^2}
  {}
  \frac{q^2(q^{2N}-1)}{q^2-1}
\end{align}
where for simplicity we have chosen equal gauge couplings for all
$U(1)$s, $g_i \equiv g$. 

If the current $J_\mu$ is anomalous under the PQ symmetry of an
axion, then it generates a large coupling of the axion with the dark
photon, 
\begin{align}
k \alpha_D \simeq \frac{g_D^2}{4\pi} q^{2N} \simeq \frac{g^2}{4\pi}
\,.
\end{align}

To generate a large kinetic mixing, we can add fermions which are
charged under $A_1$ as well as under the SM hypercharge with
$\mathcal{O}(1)$ charges. These will have a charge $(1,q^N)$ under
$(U(1)_Y,U(1)_D)$, as required for enhanced kinetic mixing.
Alternatively, if the SM $(B-L)$ current couples to $A_N$, then the SM
fields are coupled to the dark photon with unit charge and extremely
small gauge coupling.

Of course, we still need to invoke a mild form of additional alignment
for the axion in order to achieve $k j \alpha_D \sim  \beta \simeq 100$. 

\section{Self-interactions weaken superradiance bounds}

The bounds on dark photons from superradiance may be absent if there is
significant self-interaction of the dark photon
cloud~\cite{Arvanitaki:2009fg,Arvanitaki:2010sy,Baryakhtar:2017ngi}. 
We expect that self-interactions become large when the size of the
quartic interaction energy becomes of order the gravitational binding energy. For the axion, this
is roughly when $a \sim f_a$. If this happens before the axion cloud
has a chance to extract most of the black hole angular momentum, then
the superradiant instability is quenched and the bounds are evaded. 

As in the axion case, to ascertain whether self-interactions affect the superradiance bounds we first estimate the field value at which the quartic term becomes comparably important to the gravitational binding energy term:
\begin{align}
  A_\mu \sim \alpha_G \frac{\mdp }{ \sqrt{\lambda}},
\end{align}
where we expect 
\begin{equation}
\alpha_G \equiv G M_{\rm BH} \mdp \sim 1
\end{equation}
in order for superradiance to be efficient. Then, if the black hole mass is much larger than the
energy in this largest allowed cloud, then that means that the
self-interactions will take over and quench the superradiant
instability before it extracts an appreciable amount of black hole
spin. Thus, superradiance bounds only apply if,
\begin{align}
  \lambda_{\gamma'} 
  &\lesssim \left(\frac{\mdp}{M_{\rm Pl}}\right)^2,
  \label{eq:SR-self}
\end{align}
dropping factors of $\alpha_G$ and ignoring $\mathcal{O}(1)$ factors
of the black hole spin. This limit is mildly more stringent than the
limit derived for our mechanism in \eqref{eq:smallquartic}. 
There is an additional
model-independent contribution to the quartic 
induced by the PQ quarks responsible for
coupling the axion to the dark photons, but this is usually a
subdominant constraint as discussed in more detail above.

Note that the condition for a
superradiant instability to extract an order-one fraction of the black
hole's energy (assuming $\alpha_G \sim 1$) is parametrically that
$A_\mu \gtrsim M_{\rm Pl}$. Hence, if we do not consider
super-Planckian field ranges, effective superradiance will always
happen only in a narrow part of parameter space where various
order-one factors conspire to allow a slightly sub-Planckian field
range to suffice for extracting most of the black hole's energy.

\section{Lattice Simulation Details}

In our numerical lattice simulations, we use conformal time,
$\tau$, as the time variable and redefine field variables as $\varphi =
a\phi$ and ${\cal A}_0 = a A_0$. Then, the evolution equations are
rewritten as
\begin{equation}
\varphi''-\nabla^2 \varphi-\frac{a''}{a} \varphi + a^3 V_\phi +\frac{\beta}{f_a a} ({\bf A}'-\nabla{\cal A}_0)\cdot (\nabla \times {\bf A}) = 0,
\end{equation} 
\begin{equation}
{\bf A}''-\nabla^2 {\bf A} + a^2 m_{\gamma'}^2 {\bf A} + 2{\cal H} \nabla {\cal A}_0 -\frac{\beta}{f_a a } \left[ (\varphi'-{\cal H} \varphi) \nabla \times {\bf A}-\nabla \varphi \times ({\bf A}'-\nabla {\cal A}_0) \right] = 0,
\end{equation}
where a prime represents a derivative with respect to conformal
time and ${\cal H} = a'/a$.
In addition, we use the Lorenz gauge condition to evolve ${\cal A}_0$,
\begin{equation}
{\cal A}'_0 + 2 {\cal H} {\cal A}_0 - \nabla \cdot {\bf A} = 0.
\end{equation}

We adopt the staggered leap-frog method to solve the system of
equations above in a lattice space with step-size $\Delta\tau$.
Each of the variables ($\varphi,\varphi',{\bf A},{\bf A}',{\cal A}_0$)
is assigned at each grid point $(i,j,k)$.
Provided those variables at $\tau$, we first advance ${\cal A}_0$, $\varphi$ and ${\bf A}$ by a half step,
\begin{equation}
{\cal A}_0(\tau+\Delta\tau/2) = {\cal A}_0(\tau) + \frac{\Delta \tau}{2} \left[ -2{\cal H}(\tau) {\cal A}_0(\tau) + \nabla \cdot {\bf A}(\tau) \right],
\end{equation}
\begin{equation}
\varphi(\tau+\Delta\tau/2) = \varphi(\tau) +\frac{\Delta\tau}{2} \varphi'(\tau),~~
{\bf A}(\tau+\Delta\tau/2)={\bf A}(\tau)+\frac{\Delta\tau}{2}{\bf A}'(\tau).
\end{equation}

Next, we update $\varphi'$ and ${\cal A}'$ by $\varphi''$ and ${\bf A}''$ respectively using the values at $\tau+\Delta\tau/2$ as follows,
\begin{equation} 
\varphi'(\tau+\Delta\tau) = \varphi'(\tau)+\Delta\tau \left[ \nabla^2\varphi+\frac{a''}{a} \varphi-a^3 V_\phi -\frac{\beta}{f_a a} ({\bf A}'-\nabla {\cal A}_0) \cdot (\nabla \times {\bf A}) \right]_{\tau+\Delta\tau/2},
\end{equation}
\begin{equation}
{\bf A}'(\tau+\Delta\tau) = {\bf A}'(\tau)
+ \Delta\tau \left[ \nabla^2 {\bf A} - a^2 m_{\gamma'}^2 {\bf A} - 2 {\cal H}\nabla {\cal A}_0 +\frac{\beta}{f_a a}\big((\varphi'-{\cal H} \varphi) \nabla\times{\bf A} - \nabla \varphi \times ({\bf A}'-\nabla {\cal A}_0) \big) \right]_{\tau+\Delta\tau/2},
\end{equation}
where the expression in square brackets is evaluated at $\tau+\Delta\tau/2$.
To obtain the updated values, we replace $\varphi'(\tau+\Delta\tau/2)$ and ${\bf A}'(\tau+\Delta\tau/2)$ in the right-hand-side with their averaged value,
\begin{equation}
\varphi(\tau+\Delta\tau/2) = \frac{\varphi'(\tau+\Delta\tau)+\varphi'(\tau)}{2},~~{\bf A}'(\tau+\Delta\tau/2) = \frac{{\bf A}'(\tau+\Delta\tau)+{\bf A}'(\tau)}{2},
\end{equation}
and then, we obtain the following linear algebraic equations,
\begin{equation} \label{eq:S50}
\begin{pmatrix}
1 && X_1 && X_2 && X_3 \\
-X_1 && 1 && -Y_3 && Y_2 \\
-X_2 && Y_3 && 1 && -Y_1 \\
-X_3 && -Y_2 && Y_1 && 1
\end{pmatrix}
\begin{pmatrix}
\varphi'(\tau+\Delta\tau) \\ A'_1(\tau+\Delta\tau) \\ A'_2(\tau+\Delta\tau) \\ A'_3(\tau+\Delta\tau)
\end{pmatrix}
=
\begin{pmatrix}
Z_0 \\ Z_1 \\ Z_2 \\ Z_3
\end{pmatrix}
\end{equation}
where we have defined ${\bf X} = (X_1,X_2,X_3)$, ${\bf Y} = (Y_1,Y_2,Y_3)$ as follows,
\begin{equation}
{\bf X} = \frac{\Delta\tau}{2} \frac{\beta}{f_a a(\tau+\Delta\tau/2)} \nabla \times {\bf A}(\tau+\Delta\tau/2),~~{\bf Y} =  \frac{\Delta\tau}{2} \frac{\beta}{f_a a(\tau+\Delta\tau/2)} \nabla \varphi(\tau+\Delta\tau/2) 
\end{equation}
and $Z_0$ and ${\bf Z} = (Z_1,Z_2,Z_3)$ are defined by
\begin{equation}
Z_0 =  \varphi'(\tau)+\Delta\tau \left[ \nabla^2\varphi+\frac{a''}{a} \varphi-a^3 V_\phi \right]_{\tau+\Delta\tau/2} -\big[{\bf A}'(\tau)-2\nabla {\cal A}_0(\tau+\Delta\tau/2)\big] \cdot {\bf X},
\end{equation}
\begin{equation}
\begin{split}
{\bf Z} &= {\bf A}'(\tau) + \Delta\tau \left[ \nabla^2 {\bf A} - a^2 m_{\gamma'}^2 {\bf A} - 2 {\cal H}\nabla {\cal A}_0 \right]_{\tau+\Delta\tau/2} \\
&+[\varphi'(\tau)-2{\cal H}(\tau+\Delta\tau/2) \varphi(\tau+\Delta\tau/2)] {\bf X} + [{\bf A}'(\tau)-2\nabla {\cal A}_0(\tau+\Delta\tau)] \times {\bf Y}.
\end{split}
\end{equation}
One can obtain $\varphi'(\tau+\Delta\tau)$ and ${\bf A}'(\tau+\Delta\tau)$ by solving analytically the above algebraic equations (\ref{eq:S50}).
Finally, the updated values of ${\cal A}_0$, $\varphi$ and ${\bf A}$ can be obtained as follows,
\begin{equation}
\varphi(\tau+\Delta\tau) = \varphi(\tau+\Delta\tau/2) +\frac{\Delta\tau}{2} \varphi'(\tau+\Delta\tau),~~{\bf A}(\tau+\Delta\tau)={\bf A}(\tau+\Delta\tau/2)+\frac{\Delta\tau}{2}{\bf A}'(\tau+\Delta\tau)
\end{equation}
and 
\begin{equation}
{\cal A}_0(\tau+\Delta\tau) = {\cal A}_0(\tau+\Delta\tau/2) + \frac{\Delta \tau}{2} \left[ -2{\cal H}(\tau+\Delta
\tau) {\cal A}_0(\tau+\Delta\tau) + \nabla \cdot {\bf A}(\tau+\Delta\tau) \right],
\end{equation}
which leads to
\begin{equation}
{\cal A}_0(\tau+\Delta\tau) = \frac{{\cal A}_0(\tau+\Delta\tau/2) + (\Delta \tau/2) \nabla \cdot {\bf A}(\tau+\Delta\tau)}{1+\Delta\tau{\cal H}(\tau+\Delta\tau)}.
\end{equation}
One can obtain the field values at arbitrary time by repeating the above procedure.
It should be noted that the constraint equation given by the variation of the action with respect to $A_0$ is automatically satisfied if it is initially satisfied.

The exact updated values are given by $f(\tau+\Delta\tau)_{\rm exact} = \exp(\Delta\tau \partial_\tau) f(\tau)$ with $f=(\varphi,\varphi',{\bf A},{\bf A}',{\cal A}_0)$  and the above numerical formula coincides with the exact one up to ${\cal O}(\Delta\tau^2)$ terms, i.e.~$f(\tau+\Delta\tau)_{\rm numerical} = f(\tau+\Delta\tau)_{\rm exact}+{\cal O}(\Delta\tau^3)$.
The spatial derivative is replaced by the spatial differences in the numerical calculus and to keep second order accuracy, we adopt the mid-point formula, e.g.~$\partial_x \varphi(i,j,k) = (\varphi(i+1,j,k)-\varphi(i-1,j,k))/(2h)$ with $h$ being the grid interval.
The stepsize is determined by the Nyquist frequency and the axion/dark photon masses to follow the mode with the maximum wavenumber and the oscillation of the axion/dark photon, namely, $\Delta\tau = 0.25\times\min(h,\min(m_a^{-1},m^{-1}_{\gamma'})/(a(\tau))$.
In addition, the boxsize and the grid number of the lattice space are set to be $L_{\rm box} = (\pi/4)m_a^{-1}$ and $N_{\rm grid}^3 = 128^3$ respectively, and then the grid interval is $h=L_{\rm box}/N_{\rm grid} = (\pi/512)m_a^{-1}$.

\section{Dark Photon DM from Axion Decays}
Here we briefly discuss another production mechanism of dark photon DM.
The axion decays into a pair of dark photons through (\ref{eq:agg}), if $m_a > 2 m_{\gamma'}$.
In order for thus produced dark photons to explain DM, their momentum must be sufficiently red-shifted by the
time of the matter-radiation equality. This requires relatively heavy axion and dark photon masses.
In the following, we assume that both axion and dark photon
 are decoupled from the SM for simplicity so that
the axion decays into only dark photons. However, if the axion lifetime is shorter than $\sim 1$sec.,
one could couple the axion to the SM without changing the results significantly. 

The decay rate of the axion into dark photons is 
\begin{align}
\Gamma(a \to \gamma' \gamma') \simeq \frac{\beta^2}{64 \pi} \frac{m_a^3}{ f_a^2},
\end{align}
where we have approximated $m_\gamma' \ll m_a$. 
The dark photon abundance is given by
\begin{align}
\frac{\rho_{\gamma'}}{s} &\simeq  \frac{2m_{\gamma'}}{m_a} \left.\frac{\rho_a}{s}\right|_{3H = m_a},
\end{align}
or equivalently,
\begin{align}
\Omega_{\gamma'} h^2 \simeq 0.05\,\theta^2
 \left(\frac{g_*}{100}\right)^{-\frac{1}{4}} 
 \left(\frac{m_{\gamma'}}{{\rm MeV}}\right)
  \left(\frac{m_a}{10 {\rm\,MeV}}\right)^{-\frac{1}{2}}
 \left(\frac{f_a}{10^{10}{\rm\,GeV}}\right)^{2},
\end{align}
where the axion is assumed to start to oscillate in the radiation dominant Universe. 
If we want the dark photon to behave as cold dark matter, it must become non-relativistic
by $z \sim 10^5$. In other words,
\begin{align}
\Gamma(a \to \gamma' \gamma') \left(\frac{m_{\gamma'}}{m_a/2}\right)^2 > \left.H\right|_{z = 10^5},
\end{align}
where the r.h.s. represents the Hubble parameter at $z = 10^5$. 
For the above mechanism to work, the Hubble parameter during inflation must be higher
than the axion mass, $m_a < H_I \lesssim 10^{14}$\,GeV.

\begin{figure*}[tp]
\centering
\includegraphics[width=0.45\textwidth]{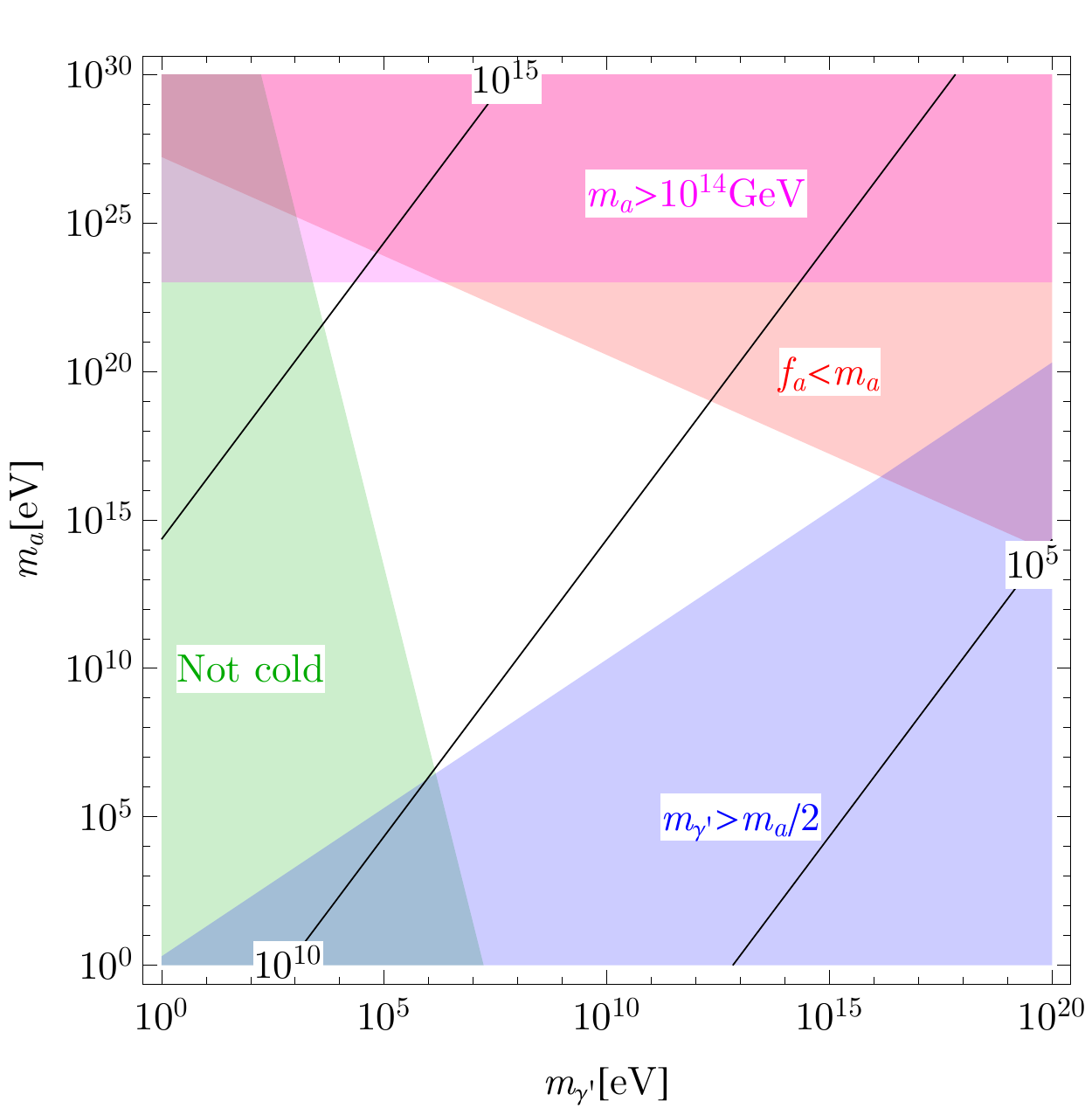}
\caption{
Dark photons produced by the axion decays can account for DM in the white region, while
the shaded regions are excluded by various requirements. See the text for details. 
The black solid lines are contours of $f_a$ in the unit of GeV.
}
\label{fig:axiondecay}
\end{figure*}

In  \cref{fig:axiondecay} we show the viable parameter space in the plane of $m_a$ and $m_{\gamma'}$,
where $f_a$ is chosen so that the dark photon abundance matches with the observed DM abundance. 
We set $\beta = 10^{-2}$ and $\theta = 1$. In the viable parameter space $m_{\gamma'}$ ranges
from keV to $10^7$\,GeV. For heavy $m_{\gamma'}$, the stability of dark photons is due to the 
absence of interactions with any lighter particles. 
If we allow a small kinetic mixing between dark photon and hypercharge
gauge boson, $m_{\gamma'}$ cannot be heavier than twice the electron mass, $m_{\gamma'} \lesssim 1$\,MeV.
In this case, the viable region is limited to $m_a \gtrsim10$\,MeV and $f_a \sim 10^{10-15}$\,GeV.

\end{document}